\documentstyle[axodraw,11pt]{article}
\def\beq{\begin{eqnarray}}    
\def\eeq{\end{eqnarray}}      

\def\de{\delta}

\def\si{\sigma}

\def\Ga{\Gamma}
\def\De{\Delta}

\begin{document}
\begin{center}

{\large\sc
The graceful exit from the anomaly-induced inflation: 
\\
Supersymmetry as a key}
\vskip 4mm

{\bf I.L. Shapiro}
 \footnote{On leave from Tomsk Pedagogical University,
Russia. E-mail: shapiro@fisica.ufjf.br},$\,$
\vskip 3mm

Departamento de F\'{\i}sica Teorica, Universidad de Zaragoza, 
50009, Zaragoza, Spain

and Departamento de F\'{\i}sica -- ICE,
Universidade Federal de Juiz de Fora, MG, Brazil
\end{center}

\vskip 6mm
\begin{quotation}

\noindent
{\large\it Abstract.}
$\,\,$ {\sl The stable version of the anomaly-induced inflation does 
not need a fine tuning and leads to sufficient expansion of the
Universe. The non-stable version (Starobinsky model) provides the
graceful exit to the FRW phase. We indicate the possibility of the
inflation which is stable at the beginning and unstable at the end.
The effect is due to the soft supersymmetry breaking and the
decoupling of the massive sparticles at low energy. }
\end{quotation}

\vskip 8mm
\noindent
{\large\bf Introduction}
\vskip 4mm

The inflation solves many problems of the Early Universe and
there are not so much doubts that it really took place (see, e.g. 
\cite{brand}). On the other hand, the conventional inflaton-based 
approach requires an exact fine-tuning of the form of the inflaton 
potential or initial data. Some people believe that there must be 
a natural mechanism for inflation, which should originate from the 
vacuum quantum effects of matter fields. The desired solution would 
not require a fine-tuning neither for initial conditions nor for 
the graceful exit to the Friedmann-Robertson-Walker (FRW) power-low 
expansion phase. The purpose of this article is to suggest a 
qualitative version of such a mechanism. Our approach is based on 
the Starobinsky model \cite{fhh,star,mamo,vile,anju} and 
supersymmetry (SUSY).  We shall also make use of the fact that SUSY 
is not seen in the low-energy phenomena. 

Consider the vacuum quantum effects in the Early Universe, when 
the typical energy of quantum processes is very high but below the 
Planck scale. Then, the appropriate framework is not the string 
theory but some quantum field theory. Furthermore, at the energies 
greater than the masses of the particles one
can apply an approximation in which the masses of the fields are
negligible. The matter filling the Universe is characterized by
pressure $\,p\,$ and energy density $\,\rho$. The standard 
relation $\,\rho=3p\,\,$ holds in the ultra-relativistic limit, 
consequently, the matter decouples from the conformal factor of the 
metric. This means local conformal (Weyl) invariance for the quantum 
fields \cite{anju}. Then, the vacuum quantum effects of the matter 
fields can be taken into account through the conformal anomaly 
\cite{fhh,mamo,star,vile,anju} and may lead to inflation \cite{star}. 
On the other hand, the masses of particles should be relevant at 
lower energies. At the intermediate scales, the 
anomaly-induced action can serve as a model for the mass-independent 
part of the effective action \cite{mamo,star,star2} while the leading 
effect of particle masses would be the renormalization of the 
Einstein-Hilbert term in the vacuum action. Taking this renormalization
into account, we arrive at the tempered form of inflation \cite{shocom}.
 
The paper as organized as follows. In the next section we 
present the qualitative description of the inflation and the graceful
exit mechanism. In section 3 some numerical estimates will be done,
in particular we establish the relation between gravitational 
scale and the radiation temperature. The last section contains 
brief general discussion of the anomaly-induced approach.

\vskip 6mm
\noindent
{\large\bf 2. Quantum effects, inflation and graceful exit}
\vskip 2mm


Consider the gauge theory which has $N_0$ real scalars,
$N_{1/2}$ Dirac spinors, and $N_1$ vectors. We shall be mainly 
interested in the vacuum effects and suppose that the interaction 
between quantum fields 
is weakened at high energies due to the asymptotic freedom. 
Than the one-loop vacuum contributions play the leading role. 
Notice that the vacuum quantum effects originate from the virtual 
particles, therefore the numbers  $\,N_{0,{1/2},1}\,$
do not describe the real matter which might fill the Universe.

The classical action of vacuum can be very complicated and 
include infinitely many local and non-local terms. However, in 
order to meet a renormalizability requirement there must be, at least,
the following three terms \cite{birdav,book}:
\beq
S_{vac} = \int d^4 x\sqrt{- g}\,
\left\{l_1C^2 + l_2E + l_3{\Box}R\,\right\}\,.
\label{vacuum}
\eeq
Here, $l_{1,2,3}$ are some parameters, $C^2$ is the square of the 
Weyl tensor and $E$ is the integrand of the Gauss-Bonnet topological 
invariant. Action (\ref{vacuum}) consists of the conformal invariant 
and surface terms. Therefore, if we add (\ref{vacuum}) to the 
Einstein-Hilbert action, then the usual (homogeneous and isotropic) 
FRW solution remains unaltered.

The renormalization of the action (\ref{vacuum}) leads to
the conformal anomaly \cite{duff,birdav,book}
\beq
<T_\mu^\mu> \,=\, - \,(wC^2 + bE + c{\Box} R)\,,
\label{anomaly}
\eeq
where $\,w,\,b,\,c\,$ are the $\beta$-functions for the
parameters $\,\l_1,\,l_2,\,l_3$
\beq
w=\frac{N_0 + 6N_{1/2} + 12N_1}{120\cdot (4\pi)^2 }\,\,,
\,\,\,\,\,\,\,
b= -\,\frac{N_0 +
11N_{1/2} + 62N_1}{360\cdot (4\pi)^2 }\,\,,
\,\,\,\,\,\,\,
c=\frac{N_0 + 6N_{1/2} - 18N_1}{180\cdot (4\pi)^2}\,.
\label{abc}
\eeq

One can consider the cosmological model directly, using the anomaly 
\cite{star}, however, it is useful to construct the anomaly-induced 
effective action. The quantum correction $\,{\bar \Ga}\,$ to 
the classical action of vacuum is related to the anomaly 
\beq
- \,\frac{2}{\sqrt{-g}}
\,g_{\mu\nu}\,\frac{\de {\bar \Ga}}{\de g_{\mu\nu}}\,=\,<T_\mu^\mu> \,.
\label{mainequation}
\eeq
The solution of this equation can be found explicitly in the form
\cite{rei,buodsh}:
\beq
{\bar \Ga} = S_c[{\bar g}_{\mu\nu}] + \int d^4 x\sqrt{-{\bar g}}\,\{
w\si {\bar C}^2 + b\si ({\bar E}-\frac23 {\bar {\Box}} {\bar R})
+ 2b\si{\bar \De}\si\,\} - \frac{3c+2b}{36}\,\int d^4 x\sqrt{-g}\,R^2\,,
\label{quantum}
\eeq
where we use, along with the original metric $g_{\mu\nu}$,
another variables: the new metric $\,{\bar g}_{\mu\nu}$ 
with the fixed determinant and \,$\si ,\,$ such that
$\,\,g_{\mu\nu} = {\bar g}_{\mu\nu}\cdot e^{2\si}.\,\,$ 
The effective action
(\ref{quantum}) includes an unknown conformal-invariant functional 
$S_c[g_{\mu\nu}]$, which is the only indefinite component of the solution 
(\ref{quantum}). This term is indeed irrelevant when we consider 
the cosmological (homogeneous and isotropic) metrics. In this case
(\ref{quantum})
is the {\it exact} one-loop correction to the classical action of 
vacuum. The second term on the {\it r.h.s.} of (\ref{quantum}) is 
non-covariant but it can be rewritten in a covariant non-local 
form \cite{rei,a} \footnote{The non-local nature of the
anomaly contribution has been first noticed in \cite{ddi}.}.

Now we are in a position to discuss the arbitrariness in the 
classical action of vacuum. It is important that this is the
action of an {\it external} gravitational field and that the metric 
is not quantized. Then, as it was already mentioned above, one can 
add to the vacuum action (\ref{vacuum}) any local or non-local terms. 
In particular, one can introduce terms similar to those which 
emerge as quantum corrections (\ref{quantum}). 
In this way, one can change the numerical coefficients
of {\it all} terms in anomaly (\ref{anomaly}) or even cancel
the anomaly completely. Of course, if we insist that the classical 
action of vacuum should be local, then only the last 
$\,\int \sqrt{-g}\Box R$-term in (\ref{anomaly}) can be modified by 
introducing the 
$\,\int \sqrt{-g}R^2$-term into the classical action. Sometimes, this
operation is called "introducing finite counterterm".

The introduction of some ($\,\int \sqrt{-g}R^2$-type or non-local)
extra terms into the classical action of gravity may be equivalent
to the introduction of the inflaton-like fields 
(see, e.g., \cite{bar,wave}).
From our point of view, the necessity to adjust the classical action
of vacuum {\it after} the calculation of quantum corrections 
means that the program of "natural" inflation fails.
Hence, we are not going to introduce any special vacuum terms. 
Neither the coefficient of the $\,\int \sqrt{-g}R^2$-term in the
classical action of vacuum will not be taken unnaturally large. 
On the contrary, we assume that this coefficient is essentially 
smaller than the one in (\ref{quantum}) and then, for
the sake of simplicity, set it to zero.

The cosmological model is based on the action
\beq
S_{total}\, =\, -\, \frac{1}{16\pi G}\,\int d^4x\,\sqrt{-g}\,R\,
+\, S_{matter} \,+\, S_{vac}\, + \,{\bar \Ga}\,,
\label{tota}
\eeq
where the quantum correction $\,{\bar \Ga}\,$ is given
by (\ref{quantum}). It proves useful to introduce the following
variables: conformal factor $\,a=e^\si$, physical time $\,t\,$
(where $\,dt=a(\eta)d\eta$ and $\,\eta\,$ is the conformal time)
and $\,H(t)= {\dot \si}(t)$.  For the conformally flat case
(similar solutions for the FRW metric with $k=\pm 1$ are
also possible, see \cite{star})
$\,{\bar g}_{\mu\nu} = \eta_{\mu\nu}$, the equation for
$\,H(t)\,$ has the form \cite{anju}
\beq
{\stackrel{...} {H}}
+ 7 {\stackrel{..} {H}}H
+ 4\,\left(3 - \frac{b}{c}\right)\,
{\stackrel{.} {H}}H^2 + 4\,{{\stackrel{.} {H}}}^2
- 4\,\frac{b}{c}\,H^4 - \frac{1}{8\pi G c}\,
\left(\,2H^2 + {\stackrel{.} {H}}\,\right)  = 0\,,
\label{logs}
\eeq
There are special solutions with $\,H = const$:
\beq
H_0=0 \,\,\,\,\,\,\,\,\,\,\,\,\,\,{\rm and}
\,\,\,\,\,\,\,\,\,\,\,\,\,\,
H_{1/2} = \pm \frac{1}{\sqrt{-16\pi G b}}
\,,\,\,\,\,\,\,\,\,\,\,\,\,\,\,
a(t) = a_0\cdot \exp {Ht}\,.
\label{inflation}
\eeq
Indeed the positive sign in the last expression corresponds 
to inflation \cite{star}. 

The next step is to study the stability of inflationary
solution under fluctuations of the conformal factor $\,a(t)$.
The analysis of stability can be performed analytically \cite{star}
or numerically \cite{anju}. Solution (\ref{inflation})
is stable under fluctuations of $\,H(t)\,$
if the particle content of the quantum theory
satisfies the condition $\,\,b/c < 0\,\,\,$  \cite{star}.
By using Eq. (\ref{abc}), we arrive at the following criterion of 
stability \cite{wave}:
\beq
N_1\, <\, \frac13\,N_{1/2}\, + \,\frac{1}{18}\, N_0\,.
\label{const}
\eeq
One can imagine that there was some string phase transition at the 
Planck scale. When all massive string modes decouple, we are left
with the ultra-relativistic matter described by field theory.
Then, if (\ref{const}) is satisfied, the inflation starts independent 
on the initial conditions. The numerical analysis of \cite{anju}
shows that the stabilization of inflation performs in a
fraction of the Planck time and the Universe needs just
about 100 Planck times (we define the Planck time as $t_P=1/M_P=G^{1/2})$
to expand into necessary 65 $\,e$-folds.  
The main problem is that
the stable inflation is eternal and no simple receipt is known
for the graceful exit to the FRW phase. The possible solutions of
this problem were discussed in Ref´s. \cite{anju,wave}
using the effective field theory approach. In
particular, we have supposed that when the typical energy 
decreases, the masses of matter particles become relevant and
the deviation of $\,a(t)\,$ from the exponential behavior
might lead to the graceful exit. To some extent, this letter is
devoted to the realization of this idea (see also \cite{shocom}).

Let us also mention Ref. \cite{hamada}. In this interesting
paper the classical action of vacuum is chosen in such a way
that the total $\,\int \sqrt{-g}R^2$-term in (\ref{tota}) is absent. 
Then the inflation is stable but it can be destabilized by metric 
fluctuations. Indeed, this kind of solution is theoretically possible,
but we will not discuss it here.

The non-stable version of the anomaly-induced inflation 
\cite{star,star2,vile,hhr} has another advantage: it easily breaks 
and one can achieve the graceful exit to the FRW phase \cite{star,vile}. 
Furthermore, there can be rapid oscillations of the conformal
factor after the breaking, potentially producing the reheating 
\cite{vile}. The effective action (\ref{tota})
includes the massive degree of freedom associated to $\sigma$. 
The decay of this mode into matter particles has been discussed in 
\cite{star,star2,vile,hhr}.
We remark that this massive mode is not a fundamental field but
instead it is a degree of freedom induced by the quantum effects
of matter fields. If some matter field decouples (say, because it
has a large mass), then it does not contribute to the massive 
mode of $\,\si\,$ anymore and the coefficients 
$\,N_0,\,N_{1/2},\,N_1\,$ change correspondingly. At this moment the 
energy of the induced mode can transfer into the real (not virtual)
matter sector through the creation of particles.
In the time scale, the last decoupling is the one of the lightest
neutrino. After that the particle content in (\ref{const}) is
$N_1=1$ (one photon) and $N_0=N_{1/2}=0$. It is easy to see
that in this case the inflationary solution is unstable while 
the solution $H_0$ in (\ref{inflation}) is stable. Due to this 
circumstance, in the present day Universe we do not have fast 
inflation! This simple example shows, by the way, the predictive 
power of the anomaly-induced inflation. The existence of more than
18 massless scalars or more than 6 Weyl fermions is completely ruled 
out by the condition (\ref{const}).

The shortcoming of the non-stable inflation is that without
some "strong measures" it does not last long enough and the
Universe does not expand sufficiently. The necessary
"strong measures" have been discussed in \cite{star,vile}. They
consist in the extremely exact fine tuning of the initial 
conditions and in the introduction of the
$\,\int \sqrt{-g}R^2$ term into the classical action
\footnote{I am very grateful to A.A. Starobinsky for explaining
to me this point.}.
 
Let us summarize.
We have two sorts of the anomaly-induced inflation: stable
and unstable. The advantage of the stable one is that it does 
not depend on the initial data. However, it remains unclear
how the inflation stops. On the contrary, the non-stable 
inflation stops immediately and one is forced to use 
the "strong measures" in order to achieve the necessary expansion 
of the Universe. Indeed, both versions do not look perfect if 
they are considered separately. The best situation would be to have 
an inflation which is stable at the beginning and unstable at the 
end. If one could switch from one to another in a natural way, 
this could be the desired explanation of inflation.

Now we come back to the condition of stability (\ref{const}),
which depends on the particle content.
The known spectrum of particles fits with the
Minimal Standard Model (MSM). Let us present a few details.
The SM includes 6 quarks, each of them has 2 chiralities and
3 colors. Hence, quarks contribute $N_{quarks}=18$.
Furthermore, there are 6 leptons.
Taking all the neutrino massive, we arrive at $N_{leptons}=6$.
The scalar sector has one Higgs doublet $N_0=4$, while the 
vector one consists of 8 gluons, $W^{\pm},\,Z$ and photon. After all, 
we have $\,\,N_0=4,\,N_{1/2}=24,\,N_1=12\,$.
Then, Eq. (\ref{const}) indicates to the non-stable inflation and
this perfectly fits with our dream to have unstable inflation
at the end. The same concerns, obviously, the present-day 
$\,N_1=1,N_0=N_{1/2}=0\,$ case considered above. 

Now, let us remember that the anomaly-induced inflation is
supposed to occur at the sub-Planck energy domain. There are
many reasons to expect that the particle spectrum at this
scale goes beyond the SM. In particular, it may happen that
the high-energy theory possesses supersymmetry. Let us, for
example, look at the particle content of the Minimal Supersymmetric
Standard Model (MSSM). This model has $\,\,N_1=12\,$ as in the SM. 
Besides the known fermions, one needs additional superpartners 
(gaugino) to all the vectors. The same concerns Higgs particles, 
which require higgsino. In total,  we have $\,N_{1/2}=32$ for the 
MSSM. Finally,
the scalar sector includes two Higgs doublets and numerous 
superpartners of the fermions: squarks and sleptons, such that
$\,N_0=104\,$. It is easy to see that
this particle content provides stability in (\ref{const}). In fact,
similar result can be expected for any realistic supersymmetric
model. The supersymmetric extension of the gauge theory
implies the replacement of any vector multiplet by the $\,N=1\,$
vector superfield. Moreover, one has to add superpartners to the
fermions in such a way that they form chiral superfields. 
Both operations increase the number of spinors and scalars in 
(\ref{const}) and we arrive at the stable inflation. 


\noindent
\begin{picture}(120,100)(0,0)
\BCirc(60,25){25}
\Photon(0,25)(35,25){2}{8}
\Vertex(35,25){2}
\Photon(85,25)(120,25){2}{8}
\Vertex(85,25){2}
\end{picture}
$\,\,\,\,\,\,\,$
\begin{picture}(120,100)(0,0)
\BCirc(60,25){25}
\Photon(0,50)(40,40){2}{8}
\Vertex(40,40){2}
\Photon(85,25)(120,25){2}{8}
\Vertex(85,25){2}
\Photon(0,5)(40,10){2}{8}
\Vertex(40,10){2}
\end{picture}
$\,\,\,\,\,\,$
\begin{picture}(120,100)(0,0)
\BCirc(60,25){25}
\Photon(0,50)(40,40){2}{8}
\Vertex(40,40){2}
\Photon(0,0)(40,10){2}{8}
\Vertex(40,10){2}
\Photon(80,40)(120,50){2}{8}
\Vertex(80,40){2}
\Photon(80,10)(120,0){2}{8}
\Vertex(80,10){2}
\end{picture}



\begin{quotation}

\noindent
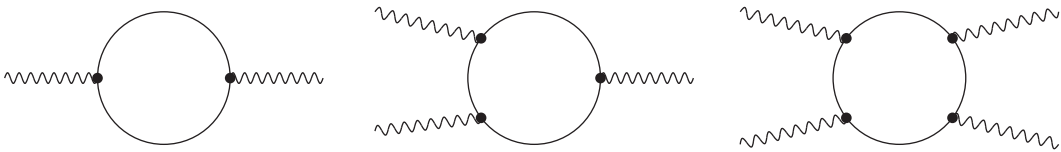
{\sl Figure.} $\,\,$
{\small \sl The samples of the one-loop Feynman diagrams which 
contribute 
to the anomaly. The bubble of the matter field has 2,3 or 4 
external lines of the field $\,\si$.}
\end{quotation}

\vskip 2mm

Thus, the situation when the anomaly-induced
inflation is stable at the beginning and unstable at the end
means exactly that supersymmetry breaks at the last stage
of inflation. If we suppose that the typical energy scale 
decreases during the inflation, this can be associated with the
supersymmetry breaking at low energies. Let us explain the
last statement in more detail. The Feynman diagrams which
contribute to the anomaly-induced action (\ref{quantum})
consist of a quantum bubble of matter (non-gravitational)
fields with external tails of the $\,\si\,$ field (see the
Figure). According to the Appelquist and Carazzone theorem
\cite{AC}, the loop of massive field decouples when the energy
of external lines becomes much smaller than the mass of the 
quantum field in the loop.
One has to notice that if the origin of masses of 
quantum fields is not the Spontaneous Symmetry Breaking,
the decoupling theorem applies nicely.
The massive sparticles decouple when the typical
energy of the external lines of the field $\,\si\,$ becomes
smaller than the masses of these particles. Therefore,
the graceful exit is realized most naturally for the soft SUSY 
breaking which is, also, the most acceptable from the
phenomenological point of view (see, e.g. \cite{susy-break}).
The mechanism of decreasing the typical energy of gravitons
during inflation will be discussed in the next section.


\vskip 6mm
\noindent
{\large\bf 3. Some numerical estimates}
\vskip 2mm


It is important to estimate the duration of inflation until the 
SUSY breaks. Since the 
breaking of the unstable inflation occurs in a very short 
time \cite{vile}, this is equivalent to the two questions:

{\it i)} What is the SUSY breaking scale?

{\it ii)}
How to evaluate the energy of the $\,\si$-field quanta in
the external lines of the diagrams?

The first
problem has been widely discussed in the literature (see, e.g.
\cite{susy-break}). The most popular is the situation when the
supersymmetry breaks softly above the electroweak
scale $\,\,M_F\approx 300\,GeV$. There are no upper bounds 
for SUSY, and its breaking may occur at the GUT scale
$M_X \approx 10^{16}\,GeV$ or even at the Planck scale 
$M_P=G^{-1/2}=10^{19}\,GeV$, depending on the SUSY model. 

To answer the {\sl ii)} question is not a simple problem. One can try 
to use the framework of renormalization group in curved space-time 
\cite{book}. This version of the renormalization group links the scaling 
of all dimensional quantities with the one of the metric. Then, as any 
dimensional quantity, the typical energy of the $\,\si$-quanta should 
vary as $\,\mu_\si \sim 1/a$, while the metric transforms as 
$g_{\mu\nu} \to g_{\mu\nu}\cdot a^2$. Unfortunately, this consideration 
is consistent only for the constant scaling parameter $\,a$. In the case 
of Eq. (\ref{inflation}) the dynamics of such important dimensional 
quantity as scalar curvature is different from the above rule, and in 
fact it is constant $\,R=-12H^2$. Since $\,H\,$ and $\,R\,$ are the 
most important local dimensional parameters associated to $\,a(t)$, 
we can identify the energy of the $a$-quanta with $\,H={\dot a}/a$. 
At low energies, when the high derivative terms in (\ref{quantum}) 
become negligible, this also follows from the Einstein equations 
which we shall apply below.

After identifying the graviton energy with $\,H\,$ we meet another 
problem. The decoupling mechanism described above works if the
energy scale is decreasing during inflation, but in the exponential 
phase (\ref{inflation}) the Hubble parameter $H(t) \equiv H_1$ does 
not decrease. However, this is true only if we do not take the 
masses of the quantum fields into account. Compared to the massless 
case, the fourth derivative anomaly-induced terms are still 
present in the effective action of massive fields. Besides, 
there are three other contributions. First of all, one meets the 
renormalization 
of the Newton constant $\,G$, and its scale dependence. In a parallel 
paper \cite{shocom} we have shown that this effect can be taken into 
account and results in the following evolution of $\,H$:
\beq
H_1 = \frac{M_P}{\sqrt{-16\pi b}}
\,\longrightarrow \, H_1(a)=
\left\{\,\frac{M^2_P}{-16\pi b}
\,-\,\frac{1}{3\,(4\pi)^2}\, \sum_{i}N_im_i^2 
\,log\Big[\frac{a(t)}{a_0}\Big]\,\right\}^{1/2}\,,
\label{shock}
\eeq
where $N_i$ and $m_i$ are multiplicity and mass of the fermion
of the specie $i$.  

Another important effect is the renormalization of the 
cosmological constant. In this paper we do not consider this 
effect which will be discussed in details elsewhere. 
It is sufficient to mention that the cosmological constant does 
not change the graceful exit mechanism. Finally, there are 
contributions given by the infinite power series in curvature. 
Unfortunately, there is no available method for calculating 
these terms in a general form. Some existing approximations may be 
misleading, in particular because 
we can not use the weak curvature limit. Let us suppose that in 
the high energy region the leading effect is the renormalization 
of $\,G\,$ and perform our analysis using Eq. (\ref{shock}). 

For simplicity we suppose that the SUSY breaks at GUT scale,
and that the gauge group and particle content are such that 
\beq
\sum_{i}\frac{1}{3(4\pi)^2}\,N_im_i^2=(10\,M_X)^2=10^{34}GeV^2
=10^{-4}M^2_P\,.
\label{huba}
\eeq
Since the numbers $N_0,N_{1/2},N_1$ are much greater than for 
the MSSM, 
the magnitude of the Hubble parameter $H_1$ will be essentially 
smaller \footnote{This is nice, because our 
semiclassical approximation (no quantum gravity, string etc) 
becomes really consistent. At lower energies $H_1$ in 
(\ref{inflation}) could be greater, but this does not matter
at all, because after the SUSY breaking the exponential solution 
gets unstable and inflation does not occur.}
than $H_1^{(MSSM)}\approx M_P$. The exact relation between $H_1$ and
$M_P$ requires fixing the particle content of the SUSY GUT. 
Without going into these details, let us suppose that 
in this situation $\,H_1=M_P/10$. The inflation will 
last until the $\,H_1(a)\,$ becomes comparable to $\,M_X$. Then
(\ref{shock}) and (\ref{huba}) immediately lead to the relation
\beq
a(H_1=M_X) \, \approx \, a_0\cdot e^{100}\,.
\label{100}
\eeq
It is easy to see that our suppositions about the spectrum of GUT 
fermions were not in favor of too much expansion, but still we arrived 
to much more than the minimal necessary 65 $e$-folds. The lower energy 
scale for the SUSY breaking can increase the number of e-folds to 
enormous extent. 

The next important problem is to link the energy of the $a(t)$-quanta 
with the temperature $T_r$ of the electromagnetic radiation. Consider 
the lower energy region after the SUSY breaks down and the inflation 
becomes unstable. Then the higher derivative terms in the effective 
action are irrelevant and the graviton energy $H$ can be defined
from the Einstein equation with the radiation dominating 
the matter stress tensor. The standard estimate \cite{brand} shows
that 
\beq
T_r \,\approx\,(H\cdot M_P)^2\,.
\label{radiation}
\eeq
Taking $H\sim M_P$ we obtain $T_r\sim M_P$ that corresponds to 
the unification of all fundamental sources and the limit of
applicability of the semiclassical approximation. Taking 
(as in our estimate leading to (\ref{100})) the value $\,H=M_X$,
we meet the temperature satisfying $\,M_X < T_r < M_P$.
In order to illustrate the gravitational suppression of $\,T_r$, 
let us consider the energy when the lightest
neutrino decouples from gravitons $H=m_\nu\approx 10^{-3}\,eV$.
According to (\ref{radiation}) the corresponding temperature
is $\,T_r\,\geq\,10^3\,GeV$. This temperature is much higher 
than the QCD and EW scales, such that all vectors of the MSM 
should be considered massless. According to (\ref{const}), the 
instability of inflation is guaranteed. An important 
consequence is that the nucleosynthesis can not be jeopardized
by the anomaly-induced inflation.

Another aspect of the energy scale problem is the following.
The change of the regime of expansion due to the decoupling of 
massive sparticles should provoke yet another quantum effects 
like particle creation. The new particles absorb the energy 
of the decaying massive mode of the $\sigma$ field, as it was 
discussed in \cite{star,vile}. Also, during the inflation 
(\ref{inflation}), the real matter filling 
the Universe is  out of the equilibrium state and does not lose 
energy. Therefore, the content of the Universe after the end 
of inflation are both hot matter which came from the string 
epoch and did not completely cool down, and the hot matter created 
during the transitional period from the stable to the unstable 
inflation. After this transition is over, the Universe entered 
into the FRW phase and all processes proceed in a standard way.

After the transition into an unstable phase, there can be, in general, 
very different versions of further behaviour for $\,a(t)\,$ 
\cite{fhh,ander,anju}. The analysis of the phase structure \cite{star} 
shows that $\,a(t)\,$ has various attractors and only some of them 
can be associated to the graceful exit to the FRW solution. The 
evolution of $\,a(t)\,$ phase can be studied numerically. The result 
depends on the initial data 
$\,a(0),{\stackrel{.} {a}}(0), {\stackrel{..} {a}}(0),
{\stackrel{...} {a}}(0)$, which are not well defined since we do 
not have the
detailed description of the decoupling. In this situation 
it is better to consider different versions of these initial data. 
For example,
if we take the purely massless case (\ref{logs}) and substitute 
the initial data $a^{(k)}(0)=H^k_1$ with $H_1$ of the MSSM) 
into the equation 
with the MSM parameters $b$ and $c$, the system goes to the
undesirable "hyperinflation" \cite{anju}. However, if we take 
the masses into account, the situation changes drastically. 
From the numerical analysis of \cite{shocom} follows that at 
the end of the stable period the evolution goes in a power-like 
manner $\,a(t)\sim t^{1/n}$. Then the initial data for the 
second MSM phase must be taken such that each derivative 
must be much smaller than the previous one. The typical time 
of the stable phase is $\,100\,t_P$. If we take,
using the Planck time units \cite{anju}
\beq
a(0)=1\,,\,\,\,\,\,\,\,\,\,\,\,
{\stackrel{.} {a}}(0)=10^{-2}\,,\,\,\,\,\,\,\,\,\,\,\,
{\stackrel{..} {a}}(0)=10^{-4}\,,\,\,\,\,\,\,\,\,\,\,\,
{\stackrel{...} {a}}(0)=10^{-6}\,,
\label{initial}
\eeq
then the asymptotic behaviour of $a(t)$ is quite similar to the 
desirable $a(t)=t^{1/2}$. The possible small changes in 
(\ref{initial}) do not modify the qualitative behaviour of $a(t)$. 
Of course, these results must be accepted with the proper
caution. Let us remind that by the end of the decoupling the 
approximation (\ref{shock}) is not safe and one has to take 
into account other contributions to the effective action. This 
can lead to the more precise definition of the initial data 
$\,a^{(i)}(0)\,$ for the unstable inflation. The subsequent 
progress in the anomaly-induced inflation requires quantitative 
description of the decoupling process (unfortunately, this is a 
difficult problem similar to the one people meet in QCD) and of 
the transition to the FRW phase. After that the metric and 
density perturbation analysis 
(see, e.g., \cite{star2,muchib,wave,hhr}) should provide useful 
information for the phenomenological investigations.

\vskip 6mm
\noindent
{\large\bf 4. Discussions}
\vskip 2mm

It is interesting to discuss the general status of the anomaly-induced 
inflation. Theoretically, the phenomena of anomalous breaking of the 
local conformal symmetry is well established (see, e.g \cite{duff}). 
The same concerns the existence of the inflationary solution 
(\ref{inflation}). Of course, there is no guarantee that this kind of 
inflation really took place in the Early Universe. However, there is 
only one way to avoid the anomaly-induced inflation: to introduce the 
classical action of vacuum in such a way that the inflation
becomes unstable at all scales. As an example we mention the 
$\sqrt{-g}R^2$-term \cite{star} which could provide inequality 
$b/c > 0$ independent on the particle content $N_0,N_{1/2},N_1$. 
Then, the inflation must have another origin. But, in this paper we have 
shown that the absence of extra vacuum terms may produce the natural 
inflation and the natural graceful exit due to the SUSY breaking. 
At this level, there are no contradictions neither with the present-day 
Universe nor with the nucleosynthesis mechanism.

Further investigations of the density and metric perturbations 
are necessary. They can brink us to 
the point when experiments and observations can show whether the 
anomaly induced inflation really took place.
\vskip 8mm

\noindent
{\bf Acknowledgments.} Author is indebted to A. Belyaev, J. Sola 
and  A.A. Starobinsky for useful discussions and to Ed. Gorbar 
for critical reading the manuscript. The grant from CNPq (Brazil) 
is gratefully acknowledged.
\vskip 6mm


\end{document}